\newcommand{\be}{\begin{equation}}
\newcommand{\ee}{\end{equation}}
\newcommand{\bea}{\begin{eqnarray}}
\newcommand{\eea}{\end{eqnarray}}
\newcommand{\bean}{\begin{eqnarray*}}

\newcommand{\eean}{\end{eqnarray*}}
\font\upright=cmu10 scaled\magstep1
\font\sans=cmss10
\newcommand{\ssf}{\sans}
\newcommand{\stroke}{\vrule height8pt width0.4pt depth-0.1pt}
\newcommand{\Z}{\hbox{\upright\rlap{\ssf Z}\kern 2.7pt {\ssf Z}}}

\newcommand{\C}{{\rlap{\rlap{C}\kern 3.8pt\stroke}\phantom{C}}}
\newcommand{\R}{\hbox{\upright\rlap{I}\kern 1.7pt R}}
\newcommand{\CP}{\C{\upright\rlap{I}\kern 1.5pt P}}
\newcommand{\PP}{\hbox{\upright\rlap{I}\kern 1.5pt P}}

\newcommand{\identity}{{\upright\rlap{1}\kern 2.0pt 1}}

\newcommand{\HH}{\mbox{\hbox{\upright\rlap{I}\kern 1.7pt H}}}

\newcommand{\fr}{\frac}

\newcommand{\pr}{\partial}

\newcommand{\ve}{\varepsilon}

\newcommand{\zb}{{\bar z}}
\input{epsf}

\documentstyle[12pt,a4wide]{article}

\newcommand{\news}{\setcounter{equation}{0}}
\begin{document}
\title{\vskip -70pt
\begin{flushright}
{\normalsize UKC/IMS/99/13} \\
{\normalsize To appear in Phys. Lett. B.} \\
\end{flushright}\vskip 50pt
{\bf \large \bf Monopoles from Rational Maps}\\[30pt]
\author{Theodora Ioannidou and Paul M. Sutcliffe\\[10pt]
\\{\normalsize  {\sl Institute of Mathematics, University of Kent at Canterbury,}}\\
{\normalsize {\sl Canterbury, CT2 7NZ, U.K.}}\\
{\normalsize{\sl Email : T.Ioannidou@ukc.ac.uk}}\\
{\normalsize{\sl Email : P.M.Sutcliffe@ukc.ac.uk}}\\}}
\date{March 1999}
\maketitle

\begin{abstract}
\noindent The moduli space of charge $k$ $SU(2)$ BPS monopoles is diffeomorphic
to the moduli space of degree $k$ rational maps between Riemann spheres.
In this note we describe a numerical algorithm to compute the monopole fields
and energy density from the rational map. The results for some symmetric examples
are presented.
\end{abstract}

\newpage
\section{Introduction}
\news\ \ \ \ \ \
The BPS monopole fields, $\Phi,A_i$, are $su(2)$-valued fields in $\R^3$
satisfying the Bogomolny equation
\be
D_i \Phi=-\fr{1}{2} \ve_{ijk}\,F_{jk}
\label{Bog}
\ee
and the boundary condition that
$\|\Phi\|^2=-\frac{1}{2}\mbox{tr}\Phi^2=1$
at infinity. The Higgs field, $\Phi$, defines a map from the two-sphere at
infinity to the two-sphere of vacua and the winding number,
 $k$, of this map is a non-negative integer which equals the magnetic charge
of the monopole (in suitable units).

The moduli space of charge $k$
monopoles, ${\cal M}_k$, is a $4k$-dimensional manifold and
it has been known for some time \cite{Do} that ${\cal M}_k$
 is diffeomorphic to the space of degree $k$ based rational
maps between Riemann spheres. More recently, a new correspondence \cite{Ja}
between ${\cal M}_k$ and the space of degree $k$ unbased
 rational maps of the Riemann sphere has been proved. In this new approach
the rational map arises as the scattering data, along half-lines from the origin,
of a linear operator constructed from the monopole fields. In this note we
describe a numerical inverse scattering algorithm to compute the monopole
fields and energy density from the rational map. The results of the algorithm
are displayed for several symmetric examples.

The algorithm involves
solving a nonlinear elliptic partial differential equation which is equivalent to the
Bogomolny equation, but for which the boundary conditions are given in terms of the
rational map. It is the ability to specify a unique solution through the rational map
boundary conditions which is the crucial feature here. It eliminates a problem which
would arise in attempting a numerical solution
of the original Bogomolny equation, in that it is not at all clear
how to specify the boundary conditions on the monopole fields so as to select a
unique monopole from the $4k$-dimensional family of solutions.

An alternative numerical method for constructing monopoles, based on the Nahm transform,
 already exists \cite{HS1} and it is perhaps useful to make a brief comparison of the
two approaches. The numerical Nahm transform requires the solution of only ordinary
differential equations and not a partial differential equation, so from that point of view
it requires less computational resources. However, the input required is Nahm data,
which involves the analytic solution of a matrix ordinary differential equation.
The general solution, for charge $k$, involves theta functions on a curve of genus
$(k-1)^2$ and this is too difficult to obtain explicitly in practice. For certain
symmetric examples \cite{HMM,HS2} the curve has symmetries which allow the solution
to be obtained on a quotient curve which is elliptic, and it is such examples
which can then be used as input for the numerical code. Some special non-elliptic
examples have been studied numerically by using approximate Nahm data \cite{Su4},
but even these cases rely on very symmetric configurations and there is currently
no method available to deal with the general case for charge greater than two.
In contrast the input for the numerical algorithm presented in this note is a rational
map and so the data is free. Any rational map can be chosen as the input data and
hence the code can be used to obtain the monopole fields for all points in
${\cal M}_k$ and not merely some special cases.

Let us briefly review the main points of \cite{Ja} to see how the rational map
arises as scattering data. First introduce polar coordinates, $r,\theta,\varphi$,
and combine the angular coordinates into the Riemann sphere parameter 
$z=e^{i\varphi}\tan(\theta/2).$
The rational map is obtained by considering Hitchin's equation
\be
(D_r-i\Phi)s=0
\label{hitchin}
\ee
for the complex vector $s$, along each radial half-line from the origin out to infinity,
 with the direction of the half-line determined by the value of $z$.
The boundary condition on $\Phi$ implies that there is a one-parameter family
of solutions $s(r)=(w_1(r),w_2(r))^t$ that are bounded as $r\rightarrow\infty$.
Define $R$ to be the ratio of these
components at the origin ie. $R=w_1(0)/w_2(0).$ 
It can then be shown \cite{Ja} that $R$ is a holomorphic function of $z$ with degree
equal to the monopole charge $k.$  Thus \hbox{$R:$ \CP$^1\mapsto$ \CP$^1$} is the
required degree $k$ rational map between Riemann spheres. 
Note that the rational map is unbased,
since a gauge transformation replaces $R$ by an $SU(2)$ M\"obius transformation
determined by the gauge transformation evaluated at the origin. Thus the 
correspondence is between a monopole
and an equivalence class of rational maps, where two maps are equivalent 
if they can be mapped into each other by a reorientation of the target Riemann
sphere.
Since the above construction does not break the $SO(3)$ rotational symmetry of $\R^3$
then a monopole which is invariant
under a subgroup $G\subset SO(3)$ will have
an associated rational map $R$ which is $G$-invariant (up to M\"obius
transformations).

\section{The Algorithm}
\news\ \ \ \ \ \
Writing the Bogomolny equation (\ref{Bog}) in terms of the coordinates $r,z,\zb$
and choosing the (complex) gauge 
\be
\Phi=-iA_r=-\frac{i}{2}H^{-1}\partial_r H, \ \ \
A_z=H^{-1}\partial_z H, \ \ \ A_\zb=0
\label{gauge}
\ee
where $H\in SL(2,\C)$ is a Hermitian matrix, results in the single equation \cite{Ja}
\be
\pr_r\left(H^{-1}\, \pr_rH\right)+\fr{(1+|z|^2)^2}{r^2}
\pr_{\bar{z}}\left(H^{-1}\,\pr_zH\right)=0.
\label{jarvis}
\ee
In \cite{Ja} it is proved that solutions of this equation are uniquely determined
by boundary conditions at large $r$ which are in one-to-one correspondence
with rational maps between Riemann spheres. This result is presented more
explicitly in \cite{IS1} where it is shown that the connection between the
rational map $R(z)$ and the matrix $H(r,z,\bar z)$ is the large $r$ 
asymptotic relation
\be 
H\sim\exp\bigg\{\frac{2r}{1+\vert R\vert^2}
\pmatrix{\vert R\vert^2-1& -2\bar R\cr
-2R & 1-\vert R\vert^2\cr}\bigg\}.
\label{bc}
\ee

The input to the algorithm is the rational map $R$, and the main task is to
compute the solution of equation (\ref{jarvis}) subject to the boundary condition
that as $r\rightarrow \infty$ the solution has the asymptotic form (\ref{bc}).

To compute solutions of the elliptic equation (\ref{jarvis}) we use a
standard heat flow approach and study the parabolic equation
\be
H^{-1}\, \pr_tH=\pr_r\left(H^{-1}\,
 \pr_rH\right)+\fr{(1+|z|^2)^2}{r^2}
\pr_{\bar{z}}\left(H^{-1}\,\pr_zH\right).
\label{hf}
\ee
As $t\rightarrow \infty$ the solutions of this equation are static, that
is $t$-independent, and hence solve the original elliptic equation.

Although the choice of coordinates, $r,z,\bar z$, has played a crucial role in
the analysis so far, these are not good coordinates in which to implement a
numerical solution. In $\R^3$ the most efficient coordinates to use for
a numerical code are Cartesian coordinates, in which equation (\ref{hf})
becomes
\be H_t= H_{ii}-\frac{2}{r^2}x_iH_i
-H_i H^{-1}H_i+\frac{i}{r}\ve_{ijk}
x_iH_j H^{-1}H_k.
\label{hfc}\ee 
Here an index $i$ denotes partial differentiation with respect to 
$x_i$, and the summation convention is used over the spatial indices.

Although we are now using Cartesian coordinates the method by which we have
obtained equation (\ref{hfc}) is reflected in the broken translational symmetry.
Equation (\ref{hfc}) can not be applied at $r=0$, but this is not a problem
since there is an additional constraint $\cite{Ja}$ that at the origin $H=I$,
 the $2\times 2$ identity matrix.

We solve equation (\ref{hfc}) using a finite difference approximation, with spatial
derivatives approximated by second order symmetric differences and the time derivative
approximated by a first order forward difference. The discretization is performed
on a regular grid, typically using $70^3$ grid points. On the boundary of the
grid the matrix $H$ is fixed to be the asymptotic expression (\ref{bc}). 
Note that it is a subtle matter to enforce the boundary condition numerically since the
elements of the matrix $H$ are unbounded as $r\rightarrow\infty.$ Thus the size of the
bounding box must be big enough to ensure that the monopole configuration sits comfortably
inside the box, but the bounding box must not be too big otherwise the unbounded growth
of the elements of $H$ will lead to numerical difficulties. In practice we find that
a bounding box with $\vert {\bf x}\vert<5$ is satisfactory for all the examples we present.

The initial conditions are taken to be 
\be 
H=\exp\bigg\{\frac{2g(r)}{1+\vert R\vert^2}
\pmatrix{\vert R\vert^2-1& -2\bar R\cr
-2R & 1-\vert R\vert^2\cr}\bigg\}
\label{ic}
\ee
where $g(r)$ is a monotonically increasing function with $g(0)=0$
and $g(r)\sim r$ for large $r.$ Note that from the relation
(\ref{gauge}) between $H$ and $\Phi$ this initial condition gives
a Higgs field whose length, $\|\Phi\|^2$, is spherically
symmetric. There are no spherically symmetric $SU(2)$ monopoles
with $k>1$, so it is clear that the initial condition is 
not a good approximation to the final solution for charges
greater than one.

The elements of $H$ are evolved
directly in the time evolution and so the unit determinant constraint
is preserved by performing the rescaling transformation
\be
H\mapsto \frac{H}{\sqrt{\mbox{det} H}} 
\ee
after each iteration. A typical example requires around 5000 iterations
to converge to a static solution.

From the solution $H$ the monopole fields are computed from the relations
(\ref{gauge}), where derivatives are again approximated by second order symmetric
finite differences. For example, the length of the Higgs field, which is a gauge invariant
quantity, is given by
\be
\|\Phi\|^2=-\frac{1}{8}\mbox{tr}(\partial_r H 
\partial_r H^{-1}).
\label{higgs}
\ee
The energy density, ${\cal E}$, is then obtained from the Higgs
field via the formula
\be
{\cal E}=\partial_i\partial_i\|\Phi\|^2.
\ee

In Figure 1 we display the results of our algorithm, in the form of energy
density isosurface plots, for several examples. We have chosen some particularly
symmetric examples, with charges from two to seven, and the explicit
rational maps for each example are listed in Table 1.\\

\begin{center}
\begin{tabular}{|c|c|c|c|}
\hline
Charge & Rational Map & Symmetry & Figure\\
\hline
 & & & \\
2 & $z^2$ & Axial & 1a\\
 & & & \\
3 & $\frac{\sqrt{3}iz^2-1}{z^3-\sqrt{3}iz}$ & Tetrahedral & 1b\\
 & & & \\
4 & $\frac{z^4+2\sqrt{3}iz^2+1}{z^4-2\sqrt{3}iz^2+1}$ & Cubic & 1c\\
 & & & \\
5 & $\frac{z^5-5z}{1-5z^4}$ & Octahedral & 1d\\
 & & & \\
6 & $\frac{z^4+ia}{iaz^6+z^2}$ & Dihedral & 1e\\
 & & & \\
7 & $\frac{z^7-7z^5-7z^2-1}{z^7+7z^5-7z^2+1}$ & Dodecahedral & 1f\\
 & & & \\
\hline
\end{tabular}


{\bf Table 1} : Rational maps for the monopoles shown in Figure 1.
\ \\ 
\end{center}

The details of these rational maps and a discussion of their symmetries can be
found in ref.\cite{HMS}. A reason for choosing the examples with Platonic symmetry
is that these cases fall into the special class which have been computed with the
numerical Nahm transform \cite{HS1,HS2} and hence we can verify that the results
obtained from the two very different numerical approaches are in agreement.
The charge six example has only a dihedral symmetry, $D_{4d}$, and the
associated Nahm data is not known, since it is not an elliptic example. 
This case is therefore one in which the numerical Nahm transform could not
be used to construct the monopole and hence this is the first computation of this
particular monopole. The degree six rational map given in Table 1 has $D_{4d}$
symmetry for all values of the real parameter $a$, and the particular example
shown here is for $a=0.16.$ The reason for this particular choice is that
there is an ansatz to obtain approximate Skyrmions from rational maps \cite{HMS},
and within this ansatz the rational map with $a=0.16$ is the one which minimizes
the energy of the approximate Skyrmion. Of course, for monopoles all configurations
of a given charge have the same energy, but it is interesting to note that the
monopole energy density isosurface in Figure 1e is remarkably similar to the
Skyrmion baryon density isosurface shown in ref.\cite{HMS}. It has already been
observed that the known Platonic monopoles resemble the corresponding Skyrmions
but a deep understanding of this fact is not yet available. Within the
rational map approximation for Skyrmions there is an understanding of the 
relationship between the Skyrmion baryon density and the rational map \cite{HMS}
and it would be interesting if a similar relationship could be found to connect
the monopole energy density to the rational map.

\section{Outlook}
\news\ \ \ \ \ \
We have presented an algorithm which computes monopoles from rational maps
and displayed the results for some examples. In this section we comment on
some possible applications of this code. 

One of the surprising results obtained from the numerical computation
of monopoles is the discovery that a charge $k$ monopole can have more than
$k$ zeros of the Higgs field \cite{HS3,Su5}. The total number of zeros
counted with multiplicity is $k$, but some of the zeros can have a negative
multiplicity, which are known as anti-zeros.  The current numerical evidence is
obtained using the numerical Nahm transform and it is satisfying to confirm
the existence of anti-zeros with the algorithm presented in this note, since
the two numerical approaches are very different. Anti-zeros are indeed detected
in the new algorithm, as demonstrated in Figure 2. In Figure 2 we plot the length
of the Higgs field, $\|\Phi\|^2$, along the $x_3$-axis for the octahedral
5-monopole shown in Figure 1d. The solid curve displays the initial condition,
where the only Higgs zeros is at the origin, whereas the dashed curve represents
the solution at the end of the iteration. It can be seen that there are now three
zeros along this line, and hence by the octahedral symmetry there are seven in total.
The zero at the origin is an anti-zero giving a total multiplicity of five as required.
At present there is little understanding of the anti-zeros phenomenon and no
signature for their appearance is known in terms of the associated rational map.
However, now that an algorithm is available to construct a monopole from its
rational map it will be  possible to make a more extensive study and investigate
any possible connections with the topology of the space of rational maps.

In testing conjectures of electric-magnetic duality it has proved extremely
useful to use the correspondence between monopoles and rational maps \cite{SS}.
Some preliminary attempts \cite{HMS} have been made to use the rational map correspondence to 
determine which monopole configurations play a key role in this context
and perhaps the construction of the monopole fields themselves may prove useful.

In the geodesic approximation \cite{Ma1} the dynamics of $k$ slowly moving monopoles
can be approximated by geodesic motion in the $k$-monopole moduli space.
It is possible to find geodesics as fixed point sets of group actions on
the space of rational maps. By constructing the monopole fields for such families
of rational maps it would be possible to study various monopole scatterings in some
detail. 

Finally, the method introduced here could be extended to construct $SU(N)$
monopoles and this may be of use in understanding non-abelian clouds \cite{LWY3}.\\

\section*{Acknowledgements}
\news\ \ \ \ \ \
PMS acknowledges the EPSRC for an Advanced Fellowship and the grant GR/L88320.\\

\section*{Figure Captions}
\news\ \ \ \ \ \
Figure 1: Energy density isosurfaces for the monopoles listed in Table 1.\\

Figure 2: Plots of $\|\Phi\|^2$ along the $x_3$-axis for the 
octahedral 5-monopole. Initial condition (solid curve) and final solution
(dashed curve).\\

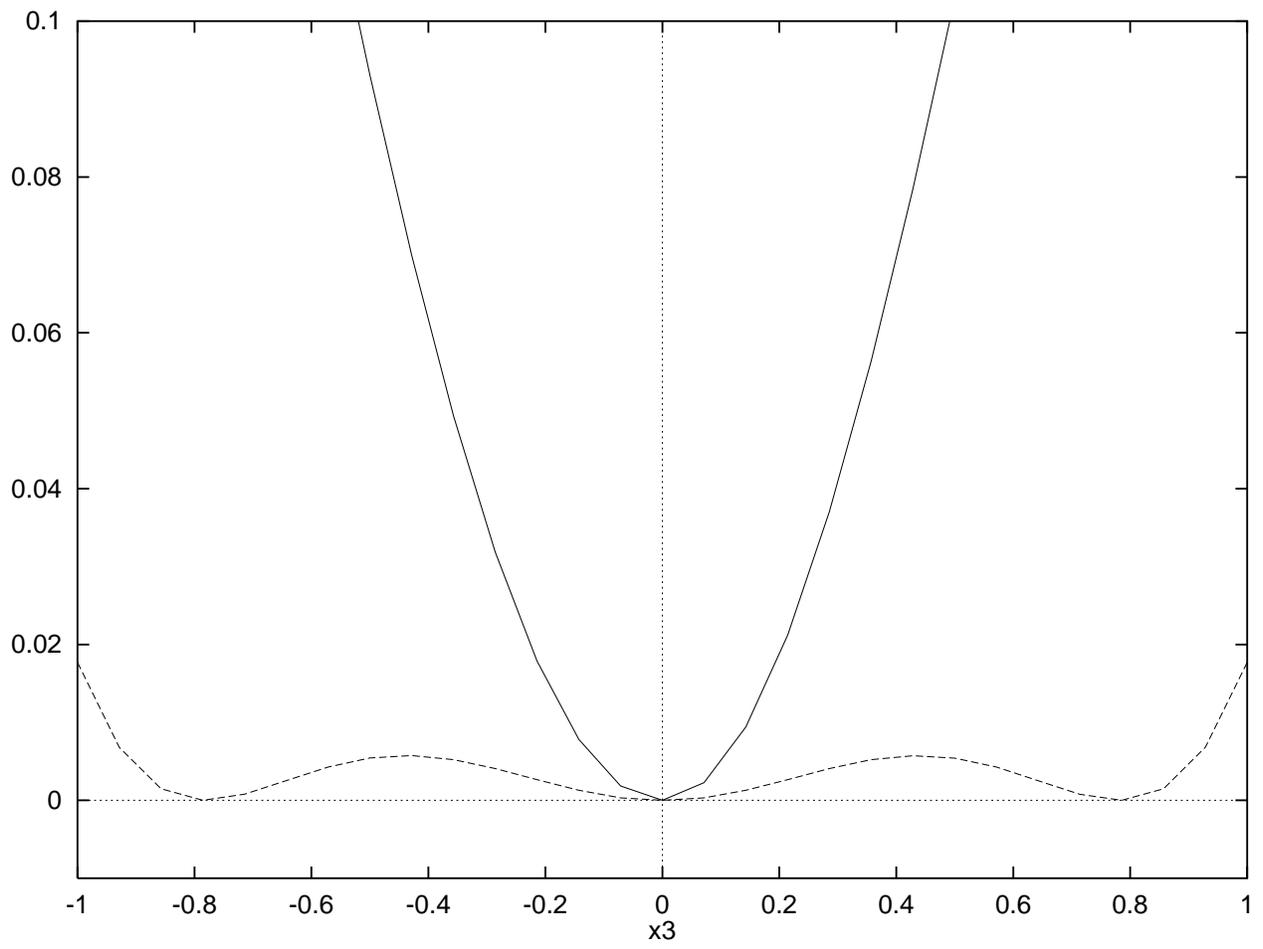
\end{document}